\documentclass{article}

    \PassOptionsToPackage{numbers, sort, compress}{natbib}
\usepackage[final]{neurips_2020}

\usepackage[utf8]{inputenc} % allow utf-8 input
\usepackage[T1]{fontenc}    % use 8-bit T1 fonts
\usepackage{hyperref}       % hyperlinks
\usepackage{url}            % simple URL typesetting
\usepackage{booktabs}       % professional-quality tables
\usepackage{amsfonts}       % blacKAoard math symbols
\usepackage{nicefrac}       % compact symbols for 1/2, etc.
\usepackage{microtype}      % microtypography

\usepackage[dvipsnames]{xcolor}

\usepackage{amsopn,amsmath,amsthm,amssymb}
\usepackage{wrapfig}
\usepackage{subcaption}
\usepackage{multirow}
 \usepackage{mathtools}

\usepackage{anyfontsize}

\usepackage{color}
\usepackage{tikz}
\usetikzlibrary{arrows,shapes,snakes,automata,backgrounds,fit,petri}
\usepackage{adjustbox}

\newcommand{\RN}[1]{%
	\textup{\lowercase\expandafter{\it \romannumeral#1}}%
}

\makeatletter
\newcommand{\distas}[1]{\mathbin{\overset{#1}{\kern\z@\sim}}}%

\usepackage{enumitem}

\usepackage{algorithm}
\usepackage{algorithmic}

\newcommand{\beq}{\vspace{0mm}\begin{equation}}
\newcommand{\eeq}{\vspace{0mm}\end{equation}}
\newcommand{\beqs}{\vspace{0mm}\begin{eqnarray}}
\newcommand{\eeqs}{\vspace{0mm}\end{eqnarray}}
\newcommand{\barr}{\begin{array}}
\newcommand{\earr}{\end{array}}

\ifx\assumption\undefined
\newtheorem{assumption}{Assumption}
\fi

\ifx\definition\undefined
\newtheorem{definition}{Definition}
\fi

\ifx\remark\undefined
\newtheorem{remark}{Remark}
\fi

\usepackage{setspace}
\usepackage{etoolbox}
\AtBeginEnvironment{quote}{\singlespacing\small}

\title{Where Is the Normative Proof? Assumptions and Contradictions in ML Fairness Research}

\author{%
  A. Feder Cooper \\
  Department of Computer Science\\
  Cornell University\\
  \texttt{afc78@cornell.edu} \\
}

\begin{document}

\maketitle

\begin{abstract}
Across machine learning (ML) sub-disciplines researchers make mathematical assumptions to facilitate proof-writing. While such assumptions are necessary for providing mathematical guarantees for how algorithms behave, they also necessarily limit the applicability of these algorithms to different problem settings. This practice is known---in fact, obvious---and accepted in ML research. However, similar attention is not paid to the \emph{normative} assumptions that ground this work. I argue such assumptions are equally as important, especially in areas of ML with clear social impact, such as fairness. This is because, similar to how mathematical assumptions constrain applicability, normative assumptions also limit algorithm applicability to certain problem domains. I show that, in existing papers published in top venues, once normative assumptions are clarified, it is often possible to get unclear or contradictory results. While the mathematical assumptions and results are sound, the implicit normative assumptions and accompanying normative results contraindicate using these methods in practical fairness applications.
\end{abstract}

\section{Introduction} \label{sec:intro}
Writing mathematical proofs requires assumptions. For example, in machine learning (ML), when writing proofs about an algorithm's properties, to simplify the math it is common to assume that the distribution we are trying to learn is convex (or strongly convex). Assumptions like this enable us to guarantee certain logical conclusions about an algorithm's behavior, such as bounds on its convergence rate. However, such assumptions can also limit an algorithm's applicability to certain learning tasks: When applying an algorithm whose theoretical basis assumes a convex distribution to a problem that is not convex, we lose the promise of the analytical guarantees that we have proven about its behavior.

Consider the following examples from Markov chain Monte Carlo (MCMC), deep learning, and distributed machine learning algorithms. Concerning MCMC, a recent paper on Poisson minibatching for Gibbs sampling on factor graphs assumes that the factor functions are all bounded ~\cite{zhang2019poisson}. This is a strong assumption, which prohibits applying the method to several problem types \cite{zhang2020tunamh}. Batch norm \cite{ioffe2015batchnorm}, a popular algorithm for accelerating deep neural network training, makes a simplifying assumption that the authors outright acknowledge to be false in reality: that the normalized inputs between two consecutive layers are Gaussian and uncorrelated. Rather than commenting on the impact of this assumption, the authors defer its investigation to future work. Lastly, the well-known Hogwild! approach to distributed machine learning makes several simplifying assumptions: a shared memory model (i.e., processes within a single computer are the mechanism for parallelization), atomic addition, convex functions that have the property of Lipschitz-continuous differentiability, and that the model weights vector being updated is sparse \cite{niu2011hogwild}. When, for example, Hogwild! is applied to a problem that is not ``sparse enough," its ability to be correct is not well-defined \cite{desa2015taming}.

The inherent, limit-inducing nature of assumptions is also well-trodden in sociotechnical literature. This quality has been discussed generally in computer systems as introducing a particular type of ``technical bias." Unlike the kind of pre-existing biases that exist in social institutions and practices, technical bias arises from making design choices or technical assumptions that constrain the problem space:
\begin{quote}
Sources of technical bias can be found in...the attempt to make human constructs amenable to computers, when we quantify the qualitative, discretize the continuous, or formalize the nonformal \cite{friedman1996bias}.
\end{quote}
This particular notion of bias from assumptions has also been long-discussed in artificial intelligence: Simplifying assumptions in the generalization language reduce the complexity of the learning problem; this introduces bias since it limits what learned model can ultimately be expressed to a set of predetermined models \cite{mitchell1980bias}.

Moreover, this formulation of bias does not just apply to strictly technical concerns. It is well-known in fields such as information science, science and technology studies, and philosophy that technology embeds values \cite{winner1980artifacts, friedman2019valuesbook, flanagan2008values, flanagan2014values}; mathematical and technical assumptions implicate normative concerns, which in turn bias the set of possible normative outcomes. These concerns are particularly relevant in disciplines with clear social impact, such as algorithmic fairness. When deploying fairness-focused methods, the results entail normative outcomes that could directly affect discrimination in the real world. Yet, the careful attention given to mathematical assumptions has not traditionally been afforded to normative ones. While rigorous mathematical proofs clearly state their assumptions, from which they build sound inferential conclusions, normative assumptions tend to be implicit and can result in normative conclusions that are unclear or contradictory.

The primary argument of this paper is that, if we take the time to clarify the normative assumptions implicit in some fairness research, the normative conclusions that follow can actually perpetuate unfairness. The mathematical proofs may be sound---a particular choice of fairness metric may even be optimized---but the implicit normative assumptions and accompanying broader normative results suggest that these methods will not ensure fairer outcomes in practical applications. My work here is preliminary and should not be taken as exhaustive. I focus on two themes extracted from publications in top venues, which demonstrate the contradictions that arise from poorly-specified normative assumptions in recent fairness research:
\begin{itemize}[noitemsep,topsep=0pt, leftmargin=.4cm]
    \item Trying to improve the quality of ML fairness research by attempting to close the gap between human and mathematical notions of fairness; this centers human perceptions of fairness, and is predicated on the implicit normative assumptions that humans are by nature fair and share a uniform understanding of fairness (Section \ref{sec:human}).
    %\item Casting fairness in an inherent trade-off with economic notions of utility; this often exclusively frames algorithmic notions of fairness purely in economic terms and sidelines correlated yet distinct issues, such as racial disparity (Section \ref{sec:econ})
    \item Attempting to optimize the ``fairness-accuracy trade-off" to get the best of both; this often implicitly assumes accuracy, unlike fairness, does not have a normative dimension, and has led to results that unfairly shift the burden of maintaining accuracy for privileged groups to marginalized groups (Section \ref{sec:tradeoff}).
\end{itemize}

Lastly, in Section \ref{sec:suggestions} I close with two concrete suggestions for more robust ML fairness research to avoid such contradictions.

\section{Misplaced Assumptions about Human Notions of Fairness} \label{sec:human}
Recent work in technical and sociotechnical venues has begun to explore human perceptions of algorithmic fairness \cite{saha2020perception,harrison2020perception,srivastava2019notions}. Most of this work focuses on ``laypeople," but there has also been some investigation into what ML practitioners deploying models in the wild should know to ensure fairness \cite{holstein2019mls}. Overall, this work tends to take an empirical approach; it surveys individuals' reactions to fairness and elicits patterns concerning which algorithmic definitions of fairness best align with human perceptions. For example, these studies look at model classification output---both real and imagined---and map different mathematical formalizations of fairness to human intuitions. The purpose is to determine if mathematical notions of fairness would be tolerable if deployed and to potentially suggest new directions for future research, namely mathematical formalizations of fairness that perhaps better reflect human perceptions (and thus perhaps would be better tolerated in practice).

To take one of these recent technical works as an example, let us consider \citet{srivastava2019notions}. The authors frame their work as a descriptive, rather than normative. They posit that fairness is an ideal that varies based on context: Depending on the societal domain in which a decision-making model is deployed, a particular mathematical formalization of fairness may be ethically preferable to others. Based on this conception of fairness, the authors advocate for the following strategy concerning how to pick an appropriate fairness metric: Instead of trying to implement multiple, mutually incompatible definitions of fairness at once \cite{kleinberg2017impossibility, friedler2016impossibility}---such that each notion holds partially---they discover and implement the most domain-appropriate fairness definition. To facilitate this process, the authors contend that it is important to consider human perceptions of fairness: ``Because algorithmic predictions ultimately impact people's lives, we argue that the most appropriate notion of algorithmic fairness is the one that reflects people's idea of fairness in the given context" \cite{srivastava2019notions}.

While the authors categorize their work as descriptive, this argument is couched in normative assumptions regarding human nature. For one, it implicitly assumes that humans are good judges of what is actually fair and can serve as a kind of ``ground truth" against which mathematical notions of fairness can be compared. If instead we assume the opposite---that humans can actually be quite unfair---would it make sense to base fairness research choices on their perceptions?

There is a rich literature spanning cognitive psychology, evolutionary psychology, economics, and game theory that discusses human fairness. Addressing this in breadth and depth is out of the scope of this short workshop paper. Instead, for brevity (and clarity), let us examine this question via an example: The ongoing debate in the US around the fairness of affirmative action policy. In brief, affirmative action is a social policy aimed at increasing the representation of historically marginalized groups in university student and workforce populations; it attempts to implement a fairer playing field by providing individuals from marginalized backgrounds with the chance to have the same opportunities as those from more privileged backgrounds.

Affirmative action has existed as official policy in the US for decades, and yet it is still unclear whether Americans generally believe that the policy aligns with its intended goal for increased fairness. It is a fairly obvious point that affirmative action is extremely contentious at this particular moment in time in the US. Some white Americans and white supremacists, who do not feel personally responsible for systemic discrimination against BIPOC\footnote{An acronym for ``Black, Indigenous, and People of Color", used particularly in the US and Canada.} populations, feel that affirmative action puts them at a disadvantage. They claim that the policy is unfair, and in fact is responsible for ``reverse discrimination" \cite{newkirk2017white, budryk2020ut}. 

The Trump administration has been sympathetic to this view; it has supported and brought lawsuits against arguably the two most hallowed institutions in US higher education, Harvard and Yale, for their use of race in college admissions~\cite{harticollis2020harvard, stratford2020yale}. While the Harvard suit principally concerns Asian-Americans, it has received support from and has arguably been co-opted by white supremacists aligned with the goal of ending affirmative action \cite{pham2018harvard}. This controversy has also in a sense found its way into algorithmic fairness literature. Cynthia Dwork uses the term ``fair affirmative action" in her work \cite{dwork2012fairness,dwork2018fairness}; she seems to be distinguishing this notion from an alternative, unfair variant. At the same time, this term is arguably redundant, since affirmative action is fundamentally about trying to promote fairer outcomes.

The fact that issues of fairness like this are controversial at all hints at a second normative assumption in \citet{srivastava2019notions}: It is possible to effectively generalize that all people have the same perception of fairness. It is clear in the context of affirmative action that this is not the case. Different people have different notions of fairness, which can be in direct opposition to each other and ultimately lead to conflicting outcomes. In short, this implicit assumption does not allow for pluralistic ideas of fairness~\cite{berlin2013crooked, flanagan2014values}, including the possibility that notions of fairness can be community-dependent. \citet{srivastava2019notions} do focus their surveys on different societal domains, but they do not really focus on specific groups of people that may be differentially affected by different domains. There is some analysis that breaks down the results by demographic attribute, such as race---a choice that is itself interesting in terms of its normative implications. It assumes that such US-defined categories of demographic difference could be salient with respect to observing differences in fairness perceptions in the domain examples they provide, even though those examples are not as controversial or community-specific in terms of outcomes as, for example, affirmative action.

We can frame this assumption as a question in terms of stakeholders: Who gets to decide what fairness means? Similar to how \citet{srivastava2019notions} posit that the choice of an appropriate mathematical fairness definition depends on societal context, one could also imagine that this choice should depend more on the impressions of the community most affected by it \cite{rawls1971theory}. Without these kinds of contextual considerations, it is unclear if it is fair to let (generalized) human perceptions of fairness guide which fairness metric should be deployed.

\section{Emergent Unfairness in Fairness-Accuracy Trade-Off Research} \label{sec:tradeoff}

Let us now turn our attention to the second theme concerning contradictory normative results, which I have located in research related to the ``fairness-accuracy trade-off." There is a long-discussed trade-off in algorithmic fairness research concerning an inherent tension between accuracy and fairness, regardless of which metrics are used in particular to formalize accuracy and fairness. Much of this work frames fairness coming at a ``cost" in accuracy---that when we choose to create fairer models, this necessitates a penalty in how correct those models can be.\footnote{There are interesting implicit normative assumptions in this line of work, especially when such work explicitly simplifies fairness to be economic fairness. I leave this discussion to future work.} 

There has been a variety of recent work that addresses this trade-off \cite{dutta2020tradeoff,bakker2020afa,menon2018cost}. For this analysis, let us examine \citet{dutta2020tradeoff}. The authors claim that the fairness-accuracy trade-off is not inherent --- that under specific ideal conditions the trade-off simply does not exist. To articulate this point, they discuss a binary classification setting that uses Hardt's equal opportunity formalization of fairness \cite{hardt2016equality} and Chernoff information, a tool from information theory, as the measure of classification accuracy. Using these definitions, the authors show analytically that, for certain ideal distributions, it is possible to train a classifier that optimizes both fairness and accuracy. However, when not dealing with the case of ideal distributions (i.e. when the dataset exhibits bias), that same classifier does indeed exhibit the trade-off between accuracy and fairness.

\citet{dutta2020tradeoff} use \citet{friedler2016impossibility}'s formulation for distinguishing between construct space and observed space to account for this difference in classification outcomes. In short, the construct space represents the features we actually want to measure but cannot observe (e.g., intelligence). Instead, we see features in the observed space (e.g., SAT score), and there is a mapping from features in the construct space to features in the observed space (e.g., SAT score is the mapped feature in the observed space, standing in place for intelligence in the construct space). In contrast to the ideal case they discuss, \citet{dutta2020tradeoff} argue that the trade-off between classification accuracy and fairness exists in the real world due to ``noisier mappings" for less privileged groups from the construct space to the observed space. They contend that this noise comes from historic differences, particularly in opportunity and representation, which makes positive and negative labels less separable (in comparison to privileged groups) for the learned classifier. It is this decrease in separability that leads to less fair classification for less privileged groups. 

Beyond making this observation, the authors want to provide a concrete suggestion for alleviating the bias from noisy mappings when deploying classifiers in practical applications. They suggest (and analytically support) that collecting more features for the unprivileged group will help ensure fairer outcomes. That is, additional features help make the mappings from construct to observed space less noisy: The features can make the real dataset's distribution better resemble the ideal distribution, which in turn yields a classifier that behaves more similarly to the ideal case, optimizing both fairness in accuracy. In other words, the data for the unprivileged group will have improved separability, which will result in reduced bias in the classification results for that group. Moreover, the authors note that gathering more features for the unprivileged group leads to these benefits without impacting the privileged group; the privileged group's accuracy and fairness metrics remain unchanged. 

Setting aside that it might not even be possible to collect more features in practice,\footnote{To list just a few potential problems: What if additional features do not exist? What if we are using a historical dataset and are trying to retroactively make models that use those datasets fairer? What if it is too expensive to collect more data?} there are important implicit normative assumptions in this choice of solution. In particular, it seems like \citet{dutta2020tradeoff} pose data collection as a value-neutral solution to ensure greater fairness. This assumption is clearly false: Data collection, which is often a form of surveillance, is not a neutral act, and it is generally not equally applied across demographic groups in the US. 

While \citet{dutta2020tradeoff} increase their fairness metric through data collection, the choice to collect more data raises a normative question directly in contradiction with their goal for increased fairness for unprivileged groups: Do we really want to collect more data on unprivileged groups---groups that already tend to be surveilled at higher rates than those with more privilege? In the US in particular there is a long history of tracking non-white individuals: Black Americans, from Martin Luther King to Black Lives Matter activists; Japanese Americans during World War II; non-white Muslims, particularly since 9/11; Latinx individuals in relation to immigration status \cite{bedoya2016surveillance, desilver2020surveillance, speri2019surveillance}. In a more global treatment of fairness, is it fair to collect more data on these populations just to ensure we are optimizing some local fairness metric? As \citet{srivastava2019notions} suggest generally, the answer to this question might depend on the deployment context. While for the case of deciding who gets a mortgage this additional data collection would almost certainly not be fair, perhaps (though not definitively) in some domains, such as healthcare, it would be more tolerable. 

To be fair, \citet{dutta2020tradeoff} are not alone in suggesting more data collection. There is a growing subfield called ``active fairness" \cite{bakker2020afa, noriega2019active}, which aims to collect higher quality features such that accuracy can be maximized while also ensuring a particular fairness metric. But in \citet{dutta2020tradeoff}, there is an emphasis on collecting additional features \emph{only} for the unprivileged group; they leave the data for the privileged group untouched in order to decrease the separability gap between groups. In essence, this unfairly shifts the burden of producing fair classification results to the unprivileged group, affording the additional privilege (i.e. even less relative surveillance) to the already privileged group. Put another way, their solution to the bi-objective fairness-accuracy optimization problem introduces \emph{another}, unexplored objective function---an objective function concerning the burden of surveillance, whose solution in this case causes residual unfairness for the marginalized group.

\section{Foregrounding Normative Assumptions} \label{sec:suggestions}
I close with two concrete recommendations for ML fairness research to help avoid the kinds of unintended normative conflicts discussed above. Both suggestions are two variants on a theme: ML fairness researchers should foreground the normative assumptions in their work. Doing so will lead to more sensible normative claims, which could in turn prevent undesirable, institutional failures or mistakes when fairness algorithms are deployed in practice \cite{perrow1999risk,vaughan1996challenger}.

\subsection{Making Such Assumptions Explicit}
ML algorithm researchers are accustomed to clearly stating their mathematical assumptions up front and building inferential conclusions that are guaranteed to hold as long as those assumptions are met. Researchers should similarly take the time to make explicit the normative assumptions underlying their work. They should consider engaging the assistance of social scientists if they believe they lack the expertise to do so independently. Even if researchers are unable to come up with an exhaustive list of normative assumptions, the process of clarifying them could help prevent muddled normative conclusions, which in turn would make fairness algorithms more suitable for use in practice.

\subsection{Tweaking Normative Assumptions for Robustness}
Making normative assumptions explicit could help facilitate more robust ML fairness research. When investigating algorithmic robustness, researchers are comfortable with relaxing or changing certain mathematical proof assumptions and reasoning out the resulting changes (or stasis) in algorithm behavior. As the economist Edward Leamer notes:
\begin{quote}
...an inference is not believable if it is fragile, if it can be reversed by minor changes in assumptions. … A researcher has to decide which assumptions or which sets of alternative assumptions are worth reporting.
\end{quote}
As a test of normative robustness, I similarly recommend that ML researchers perturb their normative assumptions and investigate how this may alter normative outcomes. For example, when considering surveillance as an appropriate solution to the accuracy-fairness trade-off, it would be useful to state this as an explicit assumption, and then consider surveillance as a constraint for the problem. In other words, one could ask, how much surveillance is tolerable for increased fairness? Perhaps none, but perhaps there is small set of high quality features that could be collected to serve this purpose, rather than just indiscriminately collecting a lot of additional features. Overall, this practice of tweaking assumptions could help avoid ``whimsical" or ``fragile" \cite{leamer1983econ} aspects of inference regarding normative considerations, leading to more credible results.

\section*{Acknowledgments}
This work was made possible by generous funding by a Cornell Institute for the Social Sciences (ISS) Team Grant, ``Algorithms, Big Data, and Inequality" (PIs Professors Ifeoma Ajunwa and Marty Wells). I would like to thank the following individuals for feedback on earlier versions of this work: Bilan A.H. Ali, Harry Auster, Kate Donahue, Jessica Zosa Forde, and Alec Pollak.

\bibliography{references}
\bibliographystyle{plainnat}

\end{document}